\documentclass[nofootinbib,aps,preprint,superscriptaddress]{revtex4}
\usepackage{amsmath,amssymb}
\usepackage{citesort}
\usepackage{epsfig}
\usepackage{pstricks,pst-node,pst-text,pst-3d}

\newcommand{\beqa}{\begin{eqnarray}}
\newcommand{\eeqa}{\end{eqnarray}}
\newcommand{\beq}{\begin{equation}}
\newcommand{\eeq}{\end{equation}}
\newcommand{\bal}{\begin{align}}
\newcommand{\eal}{\end{align}}

\def\gsim{\ \rlap{\raise 3pt \hbox{$>$}}{\lower 3pt \hbox{$\sim$}}\ }
\def\lsim{\ \rlap{\raise 3pt \hbox{$<$}}{\lower 3pt \hbox{$\sim$}}\ }

\def\Bbar{\overline{B}^0}
\def\Abar{\overline{A}}
\def\Sbar{\overline{S}}
\def\lambdaB{\bar{\lambda}}
\def\GammaB{\overline{\Gamma}}

\begin{document}

\preprint{\vbox{
\hbox{}
\hbox{TECHNION-PH-2004-27}
\hbox{hep-ph/0407002}
\hbox{September 2004} 
}}

\vspace*{48pt}

\title{On measuring $\alpha$ in $B(t)\to \rho^\pm \pi^\mp$}

\def\addtech{Department of Physics,
Technion--Israel Institute of Technology,\\
Technion City, 32000 Haifa, Israel\vspace*{6pt}}

\author{M. Gronau}\affiliation{\addtech}
\author{J. Zupan}
\affiliation{\addtech}
\affiliation{J.~Stefan Institute, Jamova 39, P.O. Box 3000,1001
Ljubljana, Slovenia\vspace*{18pt}}

\begin{abstract} \vspace*{18pt}
Defining a most economical parametrization of time-dependent 
$B\to \rho^{\pm} \pi^{\mp}$ decays, including a measurable phase $\alpha_{\rm eff}$
which equals the weak phase $\alpha$ in the limit of vanishing penguin amplitudes, we propose two
ways for determining $\alpha$ in this processes. We explain the limitation of one method, assuming
only that two relevant tree amplitudes factorize and that their relative strong phase, $\delta_t$, is 
negligible. The other method, based on broken flavor SU(3), permits a determination 
of $\alpha$ in $B^0\to \rho^{\pm}\pi^{\mp}$ in an overconstrained system using also 
rate measurements of $B^{0,+}\to K^*\pi$ and $B^{0,+} \to \rho K$. 
Current  data are shown to restrict two ratios of penguin and tree amplitudes, $r_\pm$,
to a narrow range around 0.2, and  to imply an upper bound 
$|\alpha_{\rm eff} - \alpha| < 15^\circ$. Assuming that $\delta_t$ is much smaller than 
$90^\circ$, we find $\alpha = 
(93\pm 16)^\circ$ and $(102 \pm 20)^\circ$ using BABAR and BELLE results for 
$B(t) \to \rho^{\pm}\pi^{\mp}$. Avoiding this assumption for completeness, 
we demonstrate the reduction of discrete ambiguities in $\alpha$ with increased statistics,  and show
that SU(3) breaking effects are effectively second order in $r_\pm$.
\end{abstract}

\maketitle

\section{Introduction}
A proposal made fourteen years ago to measure the CKM angle 
$\alpha$ in $B\to\pi\pi$ through an isospin analysis~\cite{Gronau:1990ka}
was followed shortly afterwards by a suggestion to apply this technique 
also to the quasi two-body decays, $B\to \rho \pi$~\cite{Lipkin:1991st, Gronau:1991dq}. 
The analysis requires a construction of 
two pentagons, for $B$ and $\bar B$, the sides of which describe decay 
amplitudes into different-charge $\rho\pi$ final states. This is a challenging task,  
requiring precise measurements of decay rates and asymmetries in all five modes. 
In addition it also involves a large number of discrete ambiguities and in certain 
cases continuous ambiguities in $\alpha$ \cite{Gronau:1991dq}. 
A simplification occurs when the decay amplitude of $B^0\to \rho^0\pi^0$ is much
smaller than the amplitudes of the other four processes, in which case the 
pentagons turn into approximate quadrangles. Recently the BELLE collaboration 
reported evidence for $B^0\to \rho^0\pi^0$ \cite{Dragic:2004tj} at a 
level implying that this simplification may not occur in practice. This seems to indicate 
that a useful measurement of $\alpha$ using the full isospin analysis may be
impractical even with super-B-factory-like luminosities \cite{Stark:2003nq}.

A complementary and more promising way of learning $\alpha$ in these 
decays is based on performing also a time-dependent Dalitz plot analysis of  $B^0 \to 
\pi^+\pi^-\pi^0$~\cite{Snyder:1993mx}. One uses the interference between 
two $\rho$ resonance bands to study the phase differences between distinct 
amplitudes contributing to the decay. This raises issues such as
the precise shapes of the tails of the Breit-Wigner functions, and
the effect of interference with other resonant and non-resonant
contributions~\cite{Deandrea:2000tf}. A complete implementation of this method
requires higher statistics than available today. 

An interesting and more modest question, which may already be studied now
using time-dependent decay measurements of  
$B^0(\Bbar)\to \rho^\pm \pi^\mp$ by the BABAR~\cite{BaBar} and BELLE~\cite{Belle} 
collaborations, is what can be learned 
about the weak phase $\alpha$ using this limited information alone. Since
these processes involve more hadronic parameters than measurable 
quantities, further assumptions are required to answer the question in a
model-independent manner. Flavor SU(3)~\cite{SU3, GHLR}, a symmetry less precise 
than isospin, provides a suitable framework for an answer. SU(3) symmetry relates 
$B^0\to \rho^\pm\pi^\mp$ to processes of the type $B\to K^*\pi$ and $B \to 
\rho K$~\cite{VPSU3}. Allowing for certain SU(3) breaking effects, which may be justified on 
theoretical grounds and tested experimentally,  improves 
the quality of such an analysis. A recent  application of broken flavor SU(3) to the 
considerably simpler case of measuring $\alpha$ in $B^0(t)\to \pi^+\pi^-$ was 
studied in~\cite{pipi}. 

In the present paper we extend the SU(3) analysis of $B^0 \to \pi^+\pi^-$ to study 
$B^0(t)\to \rho^\pm \pi^\mp$.
We also suggest an alternative approach to measure $\alpha$ in $B\to \rho^{\pm}\pi^{\mp}$, 
which does not rely on flavor SU(3), but  instead reduces the number of hadronic  parameters 
by two when assuming that tree amplitudes factorize to a very good approximation.

Several earlier studies of $\alpha$ in $B\to\rho^\pm\pi^\mp$ have been performed.
An application of flavor SU(3) to these processes has been carried out 
in~\cite{HLLW, CKMfitter}. We study a wider range of aspects,  such as consequences 
of factorization of tree amplitudes, SU(3) breaking effects, bounds on ratios of penguin-to-tree amplitudes, and the nature of discrete ambiguities in values obtained for $\alpha$. 
For completeness, we will present a proof of several bounds 
on two {\em unmeasurable quantities}, $\alpha^{\pm}_{\rm eff} - \alpha$~\cite{HLLW, CKMfitter}, 
one of which will turn out to be stronger than bounds obtained earlier. 
We will then define a {\em measurable phase} $\alpha_{\rm eff}$ which provides an 
approximate measure for $\alpha$.

An SU(3) relation between $B^0\to \rho^\pm\pi^\pm$ and $B^0\to K^{*\pm}\pi^\mp$ 
has also been discussed in~\cite{Sun}; however this work made no use of the interference between 
$B^0$-$\Bbar$ mixing and $B\to\rho\pi$ decay amplitudes which is a crucial input in our study.
Implications of $B^0(t) \to\rho^\pm \pi^\mp$ for a global SU(3) fit to all charmless $B$ decays
into pairs of a vector and a pseudoscalar meson, $B \to VP$, were studied recently in~\cite{CGLRS}. 
Our study will focus on $B \to \rho^{\pm}\pi^{\mp}$ and on their direct SU(3) 
counterparts. Our model-independent approach differs from other studies of 
$B(t)\to \rho^\pm\pi^\mp$ involving {\em a priori} calculations of decay amplitudes and 
strong phases based on QCD and factorization~\cite{factor, BN}. 

The paper is organized as follows. Section II provides 
definitions of decay amplitudes and expressions for time-dependent decay rates in terms 
of a minimal set of parameters describing these amplitudes. Section III defines and 
discusses the use of $\alpha_{\rm eff}$, 
a measurable that is equal to $\alpha$ in the absence of penguin amplitudes. Section 
IV considers a seemingly useful method of reducing the number of hadronic parameters
by assuming approximate factorization of tree amplitudes, pointing out its intrinsic limitation.
Section V draws SU(3) relations between 
$B\to\rho^{\pm}\pi^{\mp}$ and several processes of the type $B\to K^*\pi$ and $B\to \rho K$.
In Section VI we summarize the current experimental measurements of relevant rates and 
asymmetries, deriving numerical bounds on ratios of penguin and tree amplitudes in $B^0\to \rho^{\pm}\pi^{\mp}$ and on the shift $\alpha_{\rm eff} - \alpha$, obtaining a range of values
for $\alpha$.
In Section VII we study the sensitivity to experimental errors of the flavor SU(3) method for 
determining $\alpha$. This discussion involves certain discrete 
ambiguities, which will be discussed briefly in this Section, and will be dealt with in 
more detail in an Appendix. We conclude with a summary in Section VIII.

\section{Amplitudes and time-dependent decay rates in $B\to \rho^\pm\pi^\mp$}
We start by setting notations and conventions. $B^0$ decay amplitudes,
$A_+$ and $A_-$, are denoted by the charge of the outgoing $\rho$, and corresponding 
$\Bbar$ amplitudes into charge conjugate states are denoted by $\overline{A}_+$ 
and $\overline{A}_-$, respectively:
\beq
\begin{split}
A_+ \equiv A(B^0\to \rho^+\pi^-)~,\qquad 
& A_- \equiv A(B^0\to \rho^-\pi^+)~,\\
\overline{A}_+ \equiv A(\Bbar\to \rho^-\pi^+)~,\qquad 
& \overline{A}_- \equiv A(\Bbar \to \rho^+\pi^-)~.
\label{amplitudes}
\end{split}
\eeq
Each of the four amplitudes can be expressed in terms of two terms, a``tree" and a 
``penguin" amplitude, carrying specific CKM factors. We adopt the c-convention, in 
which the top-quark has been integrated out in the $b\to d$ penguin transition and 
unitarity of the CKM matrix has  been used to move a $V^*_{ub}V_{ud}$ term into 
the tree amplitude. Absorbing absolute magnitudes of CKM factors in tree ($t$) and 
penguin ($p$) amplitudes, we write
\beq
\begin{split}
A_\pm=&e^{i \gamma} t_{\pm}+ p_{\pm}~,\\
\overline{A}_\pm=&e^{- i\gamma} t_{\pm}+ p_{\pm}~. \label{Apm}
\end{split}
\eeq
While dependence on the weak phase $\gamma$ is displayed explicitly, strong phases
are implicit in the definitions of complex amplitudes. We define three strong phase 
differences,
\beq\label{delta}
\delta_\pm=\arg\big(p_\pm/t_\pm\big)~,\qquad \delta_t=\arg\big(t_-/t_+\big)~.
\eeq
For convenience we also define ratios of penguin and tree amplitudes in the 
two processes and a ratio of the two tree amplitudes,
\beq
r_\pm \equiv \left|\frac{p_\pm}{t_\pm}\right|~,~~~~~~r_t \equiv \left|\frac{t_-}{t_+}\right|~.
\eeq

\begin{figure}
\begin{center}
\epsfig{file=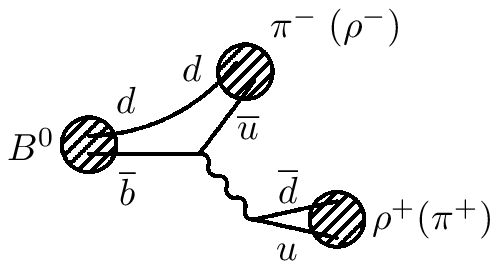}
\epsfig{file=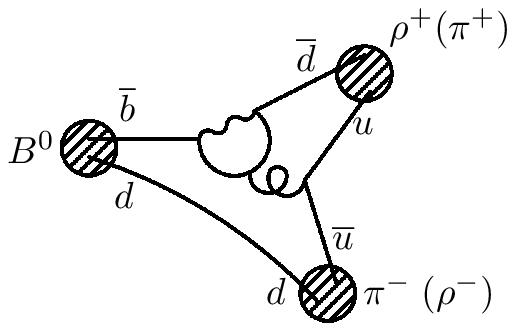}
\caption{\footnotesize{The tree (left) and penguin (right) diagrams
for the $B^0\to \rho^+\pi^-$ ($B^0\to \rho^-\pi^+$) decays.}} \label{figDecay}
\end{center}
\end{figure}

Counting parameters, we find a total of eight, consisting of seven hadronic quantities 
$|t_{\pm}|,~|p_{\pm}|, \delta_{\pm},~\delta_t$ and the weak phase $\gamma$ or $\alpha$.
We will assume $\beta$ to be given~\cite{CKMfitter, psiKs} and use $\gamma = \pi - \beta 
-\alpha$. In general, the amplitudes 
$t_+ (p_+)$ and $t_-(p_-)$ have different dynamical origins and are expected
to involve different magnitudes and different strong phases. Amplitudes with 
subscripts $+(-)$ in Eq.~\eqref{Apm} describe  transitions in which the final-state meson incorporating 
the spectator quark is a $\pi$ ($\rho$) (cf. Fig.~\ref{figDecay}). This characterization was shown to be useful in
the context of an SU(3) analysis of charmless $B$ decays into a vector and 
a pseudoscalar meson, $B\to VP$~\cite{VPSU3,CGLRS}, where $t_+ (p_+)$ and 
$t_-(p_-)$ represent SU(3) amplitudes (denoted $t_P (p_P)$ and $t_V(p_V)$ 
in~\cite{VPSU3,CGLRS}). This broader framework will be used in our discussion below.

Let us now consider measurables in time-dependent rates. Neglecting the width difference 
in the $B^0$ system, and neglecting tiny effects of CP violation in $B^0$-$\Bbar$ mixing, 
time-dependent decay rates for initially $B^0$ decaying into 
$\rho^\pm\pi^\mp$ are given by~\cite{MGPLB}
\beq\label{Gammat}
\Gamma(B^0(t) \to \rho^\pm\pi^\mp) = e^{-\Gamma t} \frac{1}{2}\left (
|A_{\pm}|^2 + |\overline{A}_{\mp}|^2\right )\left [ 1 + (C \pm \Delta C)\cos\Delta mt 
- (S \pm \Delta S)\sin\Delta mt\right ]~,
\eeq
where 
\beq\label{CSdef}
C \pm \Delta C \equiv \frac{|A_{\pm}|^2 - |\overline{A}_{\mp}|^2}{|A_{\pm}|^2 + 
|\overline{A}_{\mp}|^2}~,
~~~~~S \pm \Delta S \equiv \frac{2{\rm Im}(e^{-2i\beta}\overline{A}_{\mp}A^*_{\pm})}
{|A_{\pm}|^2 + |\overline{A}_{\mp}|^2}~.
\eeq
Here $\Gamma$ and $\Delta m$ are the average $B^0$ width and mass difference, 
respectively. For initially $\Bbar$ decays, the $\cos\Delta mt$ and $\sin\Delta mt$ 
in~(\ref{Gammat}) have opposite signs.

Counting the number of independent measurables, we find a total of six, consisting of two 
CP violating quantities, $S$ and $C$, two CP conserving measurables, 
$\Delta C$ and $\Delta S$,  and two rates, $\langle \Gamma_{\pm} \rangle \equiv 
\frac{1}{2}(|A_{\pm}|^2 + |\Abar_{\mp}|^2)$. These two rates are related to the
CP conserving charge averaged $\rho^{\pm}\pi^{\mp}$ combined decay rate, 
$\Gamma^{\rho \pi}$, and the overall CP violating asymmetry, ${\cal A}_{\rm CP}^{\rho\pi}$, 
\beq\label{Gammarhopi}
\Gamma^{\rho\pi} \equiv \langle \Gamma_+ \rangle
+ \langle \Gamma_- \rangle~,~~~
{\cal A}_{\rm CP}^{\rho\pi} \equiv \frac {\langle \Gamma_+ \rangle
- \langle \Gamma_- \rangle}
{\langle \Gamma_+ \rangle + \langle \Gamma_- \rangle}~,
\eeq
implying
\beq
\langle \Gamma_{\pm} \rangle =
\frac{1}{2}\Gamma^{\rho\pi} \left (1 \pm {\cal A}_{\rm CP}^{\rho\pi}\right )~.
\eeq
Of particular interest are the two direct CP asymmetries between $B^0(\Bbar)\to\rho^+\pi^-$ and $\Bbar(B^0)\to\rho^-\pi^+$ decay rates,
\beq\label{ACPdefined}
{\cal A}^{+}_{\rm CP} \equiv \frac{|\Abar_+|^2 - |A_+|^2}{|\Abar_+|^2 + |A_+|^2}~,~~~~~
{\cal A}^{-}_{\rm CP} \equiv \frac{|\Abar_-|^2 - |A_-|^2}{|\Abar_-|^2 + |A_-|^2}~.
\eeq
These may be expressed in terms of three of the above measurables, $C,~\Delta C$ and 
${\cal A}_{\rm CP}^{\rho\pi}$,
\beq
{\cal A}^{+}_{\rm CP} = - \frac{{\cal A}_{\rm CP}^{\rho\pi}(1 + \Delta C) + C}
{1 +  {\cal A}_{\rm CP}^{\rho\pi}C + \Delta C}~,~~~~~~~
{\cal A}^{-}_{\rm CP} = \frac{{\cal A}_{\rm CP}^{\rho\pi}(1 - \Delta C) - C}
{1 -  {\cal A}_{\rm CP}^{\rho\pi}C - \Delta C}~.
\eeq

The above observables, of which six are independent,  can be expressed in terms of the eight parameters describing $B \to \rho^\pm\pi^\mp$ in \eqref{Apm} and \eqref{delta}. This leads to rather lengthy expressions, which we do not fully display. Here we give the example of the overall CP asymmetry,  ${\cal A}_{\rm CP}^{\rho\pi}$, which does not depend on $\delta_t$,
\beqa\label{ACPrhopi}
{\cal A}_{\rm CP}^{\rho\pi}  & =  & \frac{2 \sin(\beta + \alpha)
(r_+\sin\delta_+ - r_t^2 r_-\sin\delta_-)}
{1 + r_+^2 + r_t^2 (1 + r_-^2) - 2\cos(\beta + \alpha) (r_+\cos \delta_+ + r_t^2 r_-
\cos\delta_-)}~.
\eeqa
As demonstrated by this example
which contains ratios of amplitudes rather than the amplitudes themselves, 
it is useful to consider ratios of rates, thus trading
two CP conserving measurables for two parameters $|t_+|$ and $|t_-|$. Our study 
is simplified by a judicial choice of the remaining four observables, such that they depend 
only on the six parameters $|r_\pm|$, $\delta_\pm$, $\delta_t$, and $\alpha$ without depending 
on $r_t$. 

We now display a convenient (but not unique) choice for this minimal set of observables.
Two of the observables are naturally the direct asymmetries, ${\cal A}^{\pm}_{\rm CP}$, 
which depend neither on on $r_t$ nor on $\delta_t$,
\beq\label{ACP+-}
{\cal A}^{\pm}_{\rm CP} = -\frac{2r_{\pm}\sin\delta_{\pm}\sin(\beta + \alpha)}
{1 + r^2_{\pm} - 2r_{\pm}\cos\delta_{\pm}\cos(\beta + \alpha)}~.
\eeq
Instead of $S$ and $\Delta S$ we define:
\begin{align}
\Sbar&=\frac{1}{2\sqrt{(1+\Delta C)^2
-C^2}}\left[\big(S+\Delta
S\big)\left(\frac{1+{\cal A}_{CP}^{\rho\pi}}{1-{\cal A}_{CP}^{\rho\pi}}\right)^{1/2} +\big(S-\Delta
S\big)\left(\frac{1-{\cal A}_{CP}^{\rho\pi}}{1+{\cal A}_{CP}^{\rho\pi}}\right)^{1/2} \right]~,\label{Sbar}\\
\Delta \Sbar&=\frac{1}{2\sqrt{(1+\Delta C)^2
-C^2}}\left[\big(S+\Delta
S\big)\left(\frac{1+{\cal A}_{CP}^{\rho\pi}}{1-{\cal A}_{CP}^{\rho\pi}}\right)^{1/2} -\big(S-\Delta
S\big)\left(\frac{1-{\cal A}_{CP}^{\rho\pi}}{1+{\cal A}_{CP}^{\rho\pi}}\right)^{1/2} \right]~.\label{DeltaSbar}
\end{align}
Note that $\Sbar$ and $\Delta \Sbar$ are CP violating and CP conserving, respectively, in complete analogy to $S$, $\Delta S$. They are free of  $r_t$, and their dependence on other
hadronic parameters is given by
\begin{align}
\begin{split}
\Sbar=\frac{1}{\sqrt{\quad}}&\Big\{\Big[\sin 2\alpha -(r_+ \cos\delta_+ +r_-\cos\delta_-)
\sin(\alpha-\beta)
- r_+r_-\sin 2\beta \cos(\delta_+ -\delta_-)\Big]\cos\delta_t\\
 &- \Big[(r_+ \sin\delta_+ -r_-\sin\delta_-)\sin(\alpha-\beta) + r_+r_-\sin 2\beta \sin(\delta_+ -\delta_-)\Big]\sin\delta_t\Big\}~,\label{Sbar-full}
\end{split}
\\
\begin{split}
\Delta\Sbar=\frac{1}{\sqrt{\quad}}&\Big\{\Big[\cos 2\alpha -(r_+ \cos\delta_+ +r_-\cos\delta_-)
\cos(\alpha-\beta)
+r_+r_-\cos 2\beta \cos(\delta_+ -\delta_-)\Big]\sin\delta_t\\
 &+\Big[(r_+ \sin\delta_+ -r_-\sin\delta_-)\cos(\alpha-\beta)- r_+r_-\cos 2\beta \sin(\delta_+ -\delta_-)\Big]\cos\delta_t\Big\}~,\label{DeltaSbar-full}
\end{split}
\end{align}
where 
\beq
\sqrt{\quad} \equiv \sqrt{\big(1-2r_+\cos(\beta+\alpha+\delta_+)+r_+^2\big)\big(1-2r_-\cos(\beta
+ \alpha + \delta_-)+r_-^2\big)}~.
\eeq
In our discussion in Section VII of determining $\alpha$ in a broken SU(3) analysis we will use this 
most economical parametrization of time-dependent measurements in $B^0(t)\to\rho^{\pm}\pi^{\mp}$, given by the four measurables, $A^{\pm}_{\rm CP}, \Sbar$ and $\Delta\Sbar$ in 
Eqs.~(\ref{ACP+-}), (\ref{Sbar-full}) and (\ref{DeltaSbar-full}) in terms of 
the six parameters,
$r_{\pm}, \delta_{\pm}, \delta_t$ and $\alpha$.
\section{The use of $\alpha_{\rm eff}$}\label{useofalpha}
We follow the simpler case of $B^0(t) \to \pi^+\pi^-$, where a contribution of a penguin 
amplitude modifies the value of $\alpha$ to  
$\alpha_{\rm eff}=\frac{1}{2}\arg [e^{-2i\beta}A(\Bbar \to\pi^+\pi^-)
A^*(B^0\to\pi^+\pi^-) ]$, measured from the two coefficients of 
the $\sin \Delta m t$ and $\cos\Delta mt$ terms~\cite{MGPRL}. The isospin analysis~\cite{Gronau:1990ka} 
provides a way of determining the shift $\alpha_{\rm eff} - \alpha$.
In $B\to\rho^{\pm}\pi^{\mp}$ we now define two corresponding 
quantities~\cite{BaBar,HLLW,CKMfitter},
\beq
\alpha^{\pm}_{\rm eff} \equiv \frac{1}{2}\arg\left (e^{-2i\beta}\Abar_{\pm}A^*_{\pm}\right )~.
\eeq
Note that these phases do not occur in the time-dependent rates (\ref{Gammat})
and are unmeasurable in $B\to \rho^{\pm}\pi^{\mp}$ alone. 
Instead, the observables $S \pm \Delta S$  \eqref{CSdef} involve two other related phases 
which can be measured in these decays,
\beq\label{alphaRel}
2\alpha^{\pm}_{\rm eff} \pm \hat\delta \equiv \arg\left (e^{-2i\beta}\Abar_{\pm}A^*_{\mp}\right ) = 
\arcsin\left (\frac{S\mp \Delta S}{\sqrt{1- (C\mp \Delta C)^2}}\right )~,
\eeq
where
\beq
\hat\delta \equiv \arg\left (A_-^*A_+\right )
\eeq
is an unknown relative phase between the two decay amplitudes.
Consequently, although $\alpha^{+}_{\rm eff}$ and $\alpha^{-}_{\rm eff}$ cannot be 
measured separately, their algebraic average is measurable. We therefore define:
\beq\label{alphaeff}
 \alpha_{\rm eff} \equiv \frac{1}{2}\left (\alpha^{+}_{\rm eff} + \alpha^{-}_{\rm eff} \right ) = 
\frac{1}{4}\left [
\arcsin\left (\frac{S + \Delta S}{\sqrt{1- (C + \Delta C)^2}}\right )
+ \arcsin\left (\frac{S - \Delta S}{\sqrt{1- (C - \Delta C)^2}}\right )
\right ]~.
\eeq

The two shifts $\alpha^{\pm}_{\rm eff} - \alpha$ are expected to increase with the
magnitudes of the corresponding penguin amplitudes, $|p_{\pm}|$. A relation between 
$\alpha^{\pm}_{\rm eff} - \alpha,~|p_{\pm}|,~\gamma$ and corresponding
charge averaged rates and CP asymmetries is readily obtained using Eqs.~(\ref{Apm})
(a similar relation in the different $t$-convention was shown to hold in $B\to\pi^+
\pi^-$~\cite{Charles}),
\beq\label{p^2}
4|p_{\pm}|^2\sin^2\gamma = \left (|A_{\pm}|^2 + |\Abar_{\pm}|^2 \right )
\left [ 1 - \sqrt{1 - ({\cal A}^{\pm}_{\rm CP})^2}\cos 2(\alpha^{\pm}_{\rm eff} - \alpha)
\right ]~.
\eeq
The left-hand-side of \eqref{p^2} can be bounded using flavor SU(3) as shown in 
Section V. This implies lower bounds on 
$\cos 2(\alpha^{\pm}_{\rm eff} - \alpha)$ or upper bounds on $|\alpha^{\pm}_{\rm eff} - 
\alpha|$ (see Eqs. \eqref{alphaR+}, \eqref{alphaR0} below). Using (\ref{alphaeff})  
will then provide an upper bound on $|\alpha_{\rm eff} - \alpha|$.

\section{Assuming factorization of tree amplitudes}
Since the number of parameters in $B^0(t)\to\rho^{\pm}\pi^{\mp}$ exceeds the number of 
measurables by two, a certain  input is required in order to determine $\alpha$ from these 
measurements. This input is provided by an assumption that the two tree amplitudes 
$t_{\pm}$ factorize and that their relative strong phase vanishes in this approximation. 
Given that  factorization was shown to hold to leading order in $1/m_b$ 
and $\alpha_s(m_b)$ in a heavy quark QCD expansion~\cite{BBNS, SCET}, we will
proceed under this assumption. Thus, neglecting for a moment a ratio of two  form 
factors contributing to $t_-$ and $t_+$~\cite{BN}, we take $r_t \equiv |t_-|/|t_+|$
to be given by the ratio of corresponding decay constants,
\beq\label{rt}
r_t \simeq \frac{f_\pi}{f_\rho} = 0.63~,
\eeq
where $f_\pi= 130.7~{\rm MeV}, f_\rho = 208~{\rm MeV}$.
We note that a value $r_t = 0.68$ was obtained in a global SU(3) fit to all $B \to VP$ 
decays~\cite{CGLRS}, supporting 
both factorization of tree amplitudes and 
the assumption that  $B\to \pi$ and $B\to \rho$ 
form factors do not differ much from one another.
The absolute value of $|t_+|$ obtained in the fit 
of Ref.~\cite{CGLRS} agrees with $|t_+/t| 
\simeq f_\rho/f_\pi$,
where $t$, the tree amplitude in $B^0 \to \pi^+\pi^-$, is obtained from a global SU(3)
fit to $B$ decays to two charmless pseudoscalars~\cite{CGR}. This also supports 
factorization of tree amplitudes and an assumption that the $B$ to $\pi$ form factor 
varies only slightly with $q^2$.
   
Factorization of tree amplitudes also implies that to a good approximation $\delta_t \approx 0$.
A very small phase, $\delta_t = (1\pm 3)^\circ$, supporting this assumption, 
was calculated in~\cite{BN}.  (Somewhat larger values around $-20^\circ$ were 
obtained in the global 
fit~\cite{CGLRS}.) Taking $r_t$ to be given by (\ref{rt}) and assuming $\delta_t 
\approx 0$ reduces by two the number of parameters describing $B\to\rho^{\pm}\pi^{\mp}$,  to become equal to the number of observables.
Although this situation seems perfectly suitable for a direct determination of $\alpha$,
we wish to point out its limitation.

As noted above, the four observables ${\cal A}^{\pm}_{\rm CP}, \Sbar$ and 
$\Delta\Sbar$ depend on six parameters, $r_{\pm}, \delta_{\pm}, \delta_t$ and $\alpha$, 
one of which is assumed here to vanish approximately, $\delta_t\approx 0$. The overall 
CP asymmetry 
${\cal A}^{\rho\pi}_{\rm CP}$, given explicitly in (\ref{ACPrhopi}), provides a fifth 
measurable, depending also on $r _t$, which is assumed to be given by (\ref{rt}).  
While in principle this permits a determination of $\alpha$, this can be seen to rely
on terms quadratic in $r_{\pm}$. Expanding Eqs.~(\ref{ACPrhopi}) and (\ref{ACP+-})
up to terms linear in $r_{\pm}$, we find
\beq\label{ACPrelation}
{\cal A}^{\rho\pi}_{\rm CP} = -{\cal A}^+_{\rm CP} + r_t^2{\cal A}^-_{\rm CP} + 
{\cal O}(r^2_{\pm})~.
\eeq
That is, at this order the three observables are not independent when $r_t$ is given. As 
we will show in Section VI, one expects $r_{\pm}$ to be small, $r_{\pm} \sim 0.2$, implying
that a determination of $\alpha$ using these assumptions will be very difficult.
One may turn things around, however, by using the linear relation 
(\ref{ACPrelation}) to determine $r_t$ and thereby test factorization.

\section{Constraints from flavor SU(3)}\label{constraints}
Another way of adding an input into the analysis of $B\to \rho^{\pm}\pi^{\mp}$ is provided by 
assuming flavor SU(3), as we show now. In order to improve the 
precision of our analysis, we introduce SU(3) breaking corrections in tree amplitudes.
These amplitudes, which can be shown to factorize to leading order in $1/m_b$ and 
$\alpha_s(m_b)$~\cite{BBNS, SCET}, will be assumed to involve SU(3) breaking factors 
given by ratios of meson decay constants. Penguin amplitudes, for which factorization is 
not expected to hold~\cite{SCET, charmingP}, will be assumed by default to obey exact SU(3).
The effects of SU(3) breaking in penguin amplitudes will be discussed
further in Section \ref{extracting_alpha}.
 
Strangeness changing amplitudes describing $B\to K^*\pi$ and $B\to \rho K$ will be denoted 
by primed quantities. The SU(3) counterparts of $t_+,~t_-$ and $p_{\pm}$, \eqref{Apm},
are given by~\cite{VPSU3,CGLRS}
\beqa\label{t'+}
t'_+ & = & \frac{f_{K^*}}{f_\rho}\frac{V^*_{ub}V_{us}}{V^*_{ub}V_{ud}}\,t_+ = 
\frac{f_{K^*}}{f_\rho}\lambdaB\,t_+~,\nonumber\\
\label{t'-}
t'_- & = & \frac{f_K}{f_\pi}\frac{V^*_{ub}V_{us}}{V^*_{ub}V_{ud}}\,t_- = 
\frac{f_K}{f_\pi}\lambdaB\,t_-~,\nonumber\\
\label{p'+-}
p'_{\pm} & = & \frac{V^*_{cb}V_{cs}}{V^*_{cb}V_{cd}}\,p_{\pm}
= -\lambdaB^{-1}p_{\pm}~,
\eeqa
where 
\beq
\lambdaB \equiv \frac{\lambda}{1 - \lambda^2/2} = 0.230~,~~~
\frac{f_{K^*}}{f_\rho} = 1.04~,~~~
\frac{f_K}{f_\pi} = 1.22~.
\eeq

SU(3) amplitudes represented by exchange and annihilation contributions (contributing to
$\Delta S=0$ and $\Delta S= 1$ decays, respectively) are $1/m_b$ 
suppressed relative to tree and penguin amplitudes~\cite{SCET} and will be neglected.
We also neglect very small color-suppressed electroweak penguin contributions. These 
approximations and the SU(3) breaking factors in (\ref{t'+}) can be tested in 
$B^0\to K^{*+}K^{-}, B^+\to K^{*+} \overline{K^0}$ and in other $B\to VP$ decays~\cite{CGLRS}. 
Other tests, relating CP asymmetries in $B\to \rho^\pm\pi^\mp$ and in strangeness 
changing decays, will be discussed in Section \ref{extracting_alpha}.
Under these assumptions one finds the following expressions for strangeness changing decay
amplitudes~\cite{VPSU3,CGLRS}
\beq\label{AmpPeng'}
\begin{split}
A(B^+ \to K^{*0}\pi^+)  = & -\lambdaB^{-1}p_+~,\\
A(B^+\to\rho^+K^0)  = & -\lambdaB^{-1}p_-~,
\end{split}
\eeq
and
\beq
\begin{split}\label{AmpTreePeng'}
A(B^0\to K^{*+}\pi^-) =  & \frac{f_{K^*}}{f_{\rho}}\lambdaB t_+e^{i\gamma} - 
\lambdaB^{-1}p_+~,\\
A(B^0\to \rho^-K^+) =  & \frac{f_K}{f_\pi}\lambdaB t_-e^{i\gamma} -
\lambdaB^{-1}p_-~.
\end{split}
\eeq

Denoting charge averaged decay rates by 
$\GammaB(B\to f) \equiv [\Gamma(B\to f) + \Gamma(\overline{B}\to \overline{f})]/2$,
we now define the following ratios of charge averaged rates,
\beqa\label{R*}
{\cal R}^0_+ & \equiv & \frac{\lambdaB^2\GammaB(B^0\to K^{*+}\pi^-)}
{\GammaB(B^0 \to \rho^+\pi^-)}~,~~~~~
{\cal R}^+_+  \equiv  \frac{\lambdaB^2\GammaB(B^+\to K^{*0}\pi^+)}
{\GammaB(B^0 \to \rho^+\pi^-)}~,\\
\label{R}
{\cal R}^0_- &  \equiv & \frac{\lambdaB^2\GammaB(B^0\to \rho^- K^+)}
{\GammaB(B^0\to \rho^-\pi^+)}~,~~~~~~
{\cal R}^+_-  \equiv  \frac{\lambdaB^2\GammaB(B^+\to \rho^+ K^0)}
{\GammaB(B^0\to \rho^-\pi^+)}~,
\eeqa
where superscripts and subscripts denote the charges of the $B$ and $\rho$
mesons. 
Using Eqs.~(\ref{amplitudes})-(\ref{delta}), (\ref{AmpPeng'}) and (\ref{AmpTreePeng'}),
the following expressions are obtained in terms of the hadronic parameters 
$r_{\pm}$ and $\delta_{\pm}$ and the weak phase $\beta + \alpha$:
\beq\label{R*0}
{\cal R}^0_{\pm}  =  \frac{r^2_{\pm} + 2r_{\pm}\lambdaB_{\pm}^2z_{\pm}
+ \lambdaB_{\pm}^4}{1 - 2r_{\pm}z_{\pm} +r^2_{\pm}}~,~~~~~
{\cal R}^+_{\pm} =  \frac{r^2_{\pm}}{1 - 2r_{\pm}z_{\pm} +r^2_{\pm}}~,
\eeq
where
\beq
z_{\pm} \equiv \cos\delta_{\pm}\cos(\beta + \alpha)~,~~~~~
\lambdaB_+ \equiv \sqrt{\frac{f_{K^*}}{f_\rho}}\,\lambdaB = 0.235~,~~~~~
\lambdaB_- \equiv \sqrt{\frac{f_K}{f_\pi}}\,\lambdaB = 0.254~.
\eeq

Each of these four measurables provides  an additional constraint on appropriate 
parameters. [CP asymmetries in $B^0\to K^{*+}\pi^-$ and $B^0\to \rho^-K^+$ do
not provide additional information, but can be used to test SU(3);
see Eqs. \eqref{DeltaGamma1} and \eqref{DeltaGamma2} below.] 
This leads to an overconstrained system from which $\alpha$ can 
be determined. That is, the six observables, given in 
Eqs. (\ref{ACP+-}), (\ref{Sbar-full}) and (\ref{DeltaSbar-full}) and in one pair of 
equations (\ref{R*0}), can be used to solve for the six unknowns, $r_{\pm}, 
\delta_{\pm}, \delta_t$ and $\alpha$, as discussed in more detail
 in Section VII. In the present 
section we study bounds on $r_{\pm}$ and on $\alpha_{\rm eff}- \alpha$ which 
follow from the four observables ${\cal R}^{0,+}_{\pm}$.

Each of the four expressions (\ref{R*0}) may be inverted to write $r_{\pm}$ 
in terms of $z_{\pm}$ and a corresponding ratio of rates,
\beqa\label{r+-}
r_{\pm} & = &   \frac{\sqrt{({\cal R}^0_{\pm} +\lambdaB_{\pm}^2)^2\,z_{\pm}^2 + 
(1 - {\cal R}^0_{\pm})
({\cal R}^0_{\pm} - \lambdaB_{\pm}^4)} - ({\cal R}^0_{\pm} + \lambdaB_{\pm}^2)\,z_{\pm}}
{1 - {\cal R}^0_{\pm}} 
\nonumber\\
& = & \frac{\sqrt{{\cal R}^{+2}_{\pm}\,z_{\pm}^2 + (1 - {\cal R}^+_{\pm})
{\cal R}^+_{\pm}} - {\cal R}^+_{\pm}\,z_{\pm}}{1 - {\cal R}^+_{\pm}}~.
\eeqa
The four expressions are monotonically decreasing functions of $z_{\pm}$
having their minima and maxima at $z_{\pm} = 1$ and $z_{\pm} = -1$, respectively,
\beqa\label{simple0}
\frac{\sqrt{{\cal R}^0_{\pm}}- \lambdaB^2_{\pm}}{1 + \sqrt{{\cal R}^0_{\pm}}}
\le & r_{\pm} & \le 
\frac{\sqrt{{\cal R}^0_{\pm}} + \lambdaB^2_{\pm}}{1 - \sqrt{{\cal R}^0_{\pm}}}~,\\
\label{simple+}
\frac{\sqrt{{\cal R}^+_{\pm}}}{1 + \sqrt{{\cal R}^+_{\pm}}}
\le & r_{\pm} & \le 
\frac{\sqrt{{\cal R}^+_{\pm}}}{1 - \sqrt{{\cal R}^+_{\pm}}}~.
\eeqa

Using current constraints on $\gamma$~\cite{CKMfitter}, 
$38^\circ \le \gamma \le 80^\circ$ (at $95\%$ confidence level), the lowest
and highest allowed value of $z_{\pm}$ are --0.79 and 0.79, respectively. 
This determines slightly smaller ranges of $r_{\pm}$ than given by (\ref{simple0}) 
and (\ref{simple+}) in terms of measured values of ${\cal R}^{0,+}_{\pm}$.

We note that one may use ratios of separate rates for $B$ or $\overline{B}$ mesons instead 
of the ratios of charge averaged rates defined in (\ref{R*}) and (\ref{R}).
The above considerations and the bounds on $r_{\pm}$ apply almost equally to these ratios. 
Instead of factors $z_{\pm} \equiv -\cos\delta_{\pm}\cos\gamma$ one now has factors
$\cos(\delta_{\pm} - \gamma)$ or $\cos(\delta_{\pm} + \gamma)$, which are  constrained to 
lie in a range between $-1$ to $1$. These then imply bounds of the form (\ref{simple0})
and (\ref{simple+}). For given measurements of rates and asymmetries,
as specified in the next section, one may then compare the three types of ranges obtained for 
$r_{\pm}$ and choose the most restrictive ones.

The four strangeness changing processes (\ref{AmpPeng'}) and (\ref{AmpTreePeng'}), 
which are expected to be dominated by 
penguin amplitudes, can also be used to set an upper bound on $|\alpha_{\rm eff} -\alpha|$. 
For the two charged $B$ decays one has
\beq\label{alphaR+}
\cos2(\alpha^{\pm}_{\rm eff} - \alpha) =\frac{1 - 2{\cal R}^+_{\pm}\sin^2(\beta + \alpha)}{\sqrt {1 - {\cal A}^{\pm 2}_{\rm CP} } } \ge  
\frac{1 - 2{\cal R}^+_{\pm}}{\sqrt {1 - {\cal A}^{\pm 2}_{\rm CP} } }~.
\eeq
For the two processes involving neutral $B$ decays one finds
\beqa
\frac{\lambdaB^2\GammaB(B^0\to K^{*+}\pi^-)}{|p_+|^2} & = &
1 + \lambdaB_+^4r^{-2}_+ + 2\lambdaB_+^2r^{-1}_+z_+ \ge \sin^2\gamma~,\\
\frac{\lambdaB^2\GammaB(B^0\to \rho^- K^+)}{|p_-|^2} & = &
1 + \lambdaB_-^4r^{-2}_- + 2\lambdaB_-^2r^{-1}_-z_- \ge \sin^2\gamma~,
\eeqa
where the two inequalities follow simply from the identity  and the inequality $1 + x^2 + 2x\cos\delta\cos\gamma = 1 - \cos^2\delta\cos^2\gamma + (x + \cos\delta\cos\gamma)^2
\ge \sin^2\gamma$.
Combining these inequalities with (\ref{p^2}), we find~\cite{HLLW, CKMfitter}
\beq\label{alphaR0}
\cos 2(\alpha^{\pm}_{\rm eff} - \alpha)  \ge  
\frac{1 - 2{\cal R}^0_{\pm}}{\sqrt {1 - {\cal A}^{\pm 2}_{\rm CP}}}~.
\eeq
Thus, measured branching ratios and asymmetries, appearing on the right-hand-side 
of (\ref{alphaR+}) and (\ref{alphaR0}) and listed in 
the next section, provide upper bounds on $|\alpha^{\pm}_{\rm eff} - \alpha|$ and, 
using (\ref{alphaeff}), they imply upper bounds on $|\alpha_{\rm eff} - \alpha|$.

\begin{table*}
\begin{minipage}{1.05\textwidth}
\caption{\footnotesize{Experimental charge averaged branching ratios and CP 
asymmetries  of selected $\Delta S = 0$ and $\Delta S = 1$ $B$ meson decays. 
For each process, the first line gives the branching ratio in units of $10^{-6}$, 
while the second line quotes the CP asymmetry. Note, that the averages for $\rho^+\pi^-$ and $\rho^-\pi^+$ 
final states were obtained using also the value for the summed branching ratio from CLEO and are thus not a simple
average of BaBar and Belle columns.}}
\label{table-1}
\begin{ruledtabular}
\begin{tabular}{llllll}
& Mode & CLEO & BABAR & BELLE & Avg. \\
\hline
$B^0 \to$
   & $\rho^{\pm} \pi^\mp$
       & $27.6^{+8.4}_{-7.4}\pm4.2$ \cite{Jessop:2000bv}
       & $22.6\pm1.8\pm2.2$ \cite{BaBar}
       & $29.1^{+5.0}_{-4.9}\pm4.0$ \cite{Abe:2003rj}
       & $24.0 \pm 2.5$ \\
       &
       & -
       & $-0.088\pm0.049\pm0.013$ \cite{Aubert:2004iu} 
       & $-0.16\pm{0.10}\pm 0.02$ \cite{Belle}
       & $ -0.10\pm 0.05$ \\
       & $\rho^+\pi^-$
       & -
       & $12.7\pm 2.0$
       & $19.5\pm 5.0$
       & $14.2\pm 1.9$\\
       & 
       & -
       & $-0.21 \pm 0.12 $ 
       & $-0.02\pm 0.16$ 
       & $-0.16 \pm 0.09$  \\
       & $\rho^-\pi^+$
       & -
       & $9.9\pm1.8$ 
       & $9.6\pm3.4$
       & $9.8 \pm 1.5$\\
       & 
       & -
       & $-0.47\pm{0.15}$ 
       & $-0.53\pm0.30$ 
       & $-0.48 \pm 0.14$ \\
\hline
$B^0 \to$
   & $K^{*+} \pi^-$
       & $16^{+6}_{-5}\pm2$ \cite{Eckhart:mb}
       & $11.9\pm 2.0$ \cite{Aubert:2004uf,Aubert:2004bt} 
       & $14.8^{+4.6+1.5+2.4}_{-4.4-1.0-0.9}$ \cite{Belle0317}
       & $12.7\pm 1.8$ \\ 
       &
       & $0.26^{+0.33+0.10}_{-0.34-0.08}$ \cite{Eisenstein:2003yy}
       & $-0.03\pm 0.24$ 
\footnote{The two measurements of the CP asymmetry entering this average have opposite signs, 
${\cal A}_{CP}=0.23\pm 0.18^{+0.09}_{-0.06}$ \cite{Aubert:2004uf} and ${\cal A}_{CP}=-0.25\pm 0.17\pm0.02\pm 0.02$ 
\cite{Aubert:2004bt}. The combined error includes a scaling factor ${\cal S} = 1.9$.}
 \cite{Aubert:2004uf,Aubert:2004bt} 
       & -
       & $0.06\pm 0.20$ \\ 
   & $\rho^- K^+$
       & $16.0^{+7.6}_{-6.4} \pm 2.8$ \cite{Jessop:2000bv}
       & $8.6\pm{1.4} \pm 1.0$ \cite{Aubert:2004bt} 
       & $15.1^{+3.4+1.4+2.0}_{-3.3-1.5-2.1}$ \cite{Belle0317}
       & $9.9 \pm 1.9$ \footnote{The error includes a scaling factor ${\cal S} = 1.2$.}\\ 
   &   & -
       & $0.13^{+0.14}_{-0.17}\pm 0.04 \pm 0.13$ \cite{Aubert:2004bt}
       & $0.22^{+0.22+0.06}_{-0.23-0.02}$ \cite{Belle0317}
       & $0.17 \pm 0.15$ \\
\hline
$B^+ \to$
   & $K^{*0} \pi^+$
       & $7.6^{+3.5}_{-3.0}\pm1.6$ \cite{Jessop:2000bv}
       & $10.5\pm2.0\pm1.4$ \cite{Aubert:2004fn} 
       & $9.83\pm 0.90^{+1.06}_{-1.24}$ \cite{Belle0410} 
       & $9.8\pm 1.2$ \\ 
   & $\rho^+ K^{0}$
       & $<48$ \cite{Asner:1995hc}
       & $-$
       & $-$
       & $<48$
\end{tabular}
\end{ruledtabular}
\end{minipage}
\end{table*}

\section{Current rates, asymmetries and bounds on $r_{\pm}$ and $\alpha_{\rm eff}-\alpha$}\label{currentRates}
The current measured branching ratios and asymmetries in $B^0\to \rho^{\pm}\pi^{\mp}$ 
and in SU(3) related processes are summarized in Table I
For ratios of $B^+$ and $B^0$ decay 
rates we will use the lifetime ratio~\cite{tau} $\tau(B^+)/\tau(B^0) = 1.077 \pm 0.013$.
The  BABAR~\cite{BaBar,CKMfitter,Aubert:2004iu} and BELLE~\cite{Belle} collaborations
measured in $B^0(t)\to\rho^{\pm}\pi^{\mp}$ also the four quantities,
\beq\label{CSDelta}
C = \left\{ \begin{array}{c}0.34 \pm 0.12 \cr
0.25\pm 0.17 \cr
0.31 \pm 0.10\end{array} \right.~
\Delta C= \left\{ \begin{array}{c}0.15 \pm 0.11 \cr
0.38\pm 0.18\cr
0.21\pm 0.10 \end{array} \right.~
S =\left\{ \begin{array}{c}-0.10 \pm 0.15 \cr
-0.28\pm 0.25\cr
-0.15\pm 0.13 \end{array} \right.~
\Delta S = \left\{ \begin{array}{c}0.22 \pm 0.15 \cr
-0.30\pm 0.26\cr
0.08\pm 0.23~({\cal S}=1.7),\end{array} \right.
\eeq
where the first values were obtained by BABAR, the second by BELLE and the third 
are their averages. Statistical and systematic errors were added in quadrature,
and a scaling factor ${\cal S}=1.7$ is used for the error on the averaged value of $\Delta S$.
Using these values and the definitions (\ref{Sbar}) and (\ref{DeltaSbar}), one finds
\beq\label{SbarNum}
\Sbar = \left\{ \begin{array}{c}-0.11 \pm 0.13\cr
-0.17\pm 0.18\end{array} \right.~~~~
\Delta\Sbar =  \left\{ \begin{array}{c}0.21 \pm 0.14\cr
-0.19\pm 0.19\end{array} \right.~,
\eeq
for BABAR and BELLE respectively.
In view of the difference between the values of $\Delta S$ measured by BABAR 
and BELLE, we will not only take their average in the discussion below but will also treat
them separately.

Let us now consider SU(3) bounds on the penguin pollution parameters $r_{\pm}$
 and $\alpha^{\pm}_{\rm eff} - \alpha$. Using the definitions (\ref{R*}) and (\ref{R})
and taking branching ratios from Table I, one obtains the following values: 
\beq\label{expR}
{\cal R}^0_+ = 0.048 \pm 0.010~,~~~~~{\cal R}^+_+ = 0.034 \pm 0.006~,
~~~~{\cal R}^0_- = 0.053 \pm 0.014~,
\eeq
but only an upper bound on ${\cal R}^+_-$. Applying (\ref{r+-})  and assuming Gaussian distributions, these values lead to allowed ranges for $r_{\pm}$. The most stringent bounds 
on $r_+$ follow from ${\cal R}^+_+$, which gives at 90\% confidence level:
\beq\label{r+R++}
0.14~(0.16) \le r_+ \le 0.25~(0.22)~.
\eeq
Values in parentheses are obtained by using the central value of 
${\cal R}^+_+$. The most stringent bounds on $r_-$ are obtained by using the ratios $\Gamma(B^0\to
\rho^-K^+)/\Gamma(B^0\to \rho^-\pi^+)$ and their charge conjugates, instead  of
relying on the ratio of the charge averaged rates. Using the branching ratios {\em and
asymmetries} of Table I, one finds the following range at 90\% confidence level,
\beq\label{r-bound}
0.14~(0.21) \le r_- \le 0.34~(0.29)~.
\eeq

The bounds (\ref{r+R++}) and (\ref{r-bound}) are  expected to be 
modified by additional SU(3) breaking effects
which were not  included in the analysis. In (\ref{p'+-}) we assumed exact SU(3) 
for penguin amplitudes. Somewhat smaller values of $r_{\pm}$ are 
obtained if SU(3) breaking enhances $p'_{\pm}$ relative to $p_{\pm}$, as it would, for 
instance, by assuming factorization for these amplitudes. The above bounds are 
somewhat wider than and, as expected,  consistent 
with values obtained  in a global SU(3) fit to {\em all $B\to VP$ decays}~\cite{CGLRS}, 
$r_+ = 0.17 \pm 0.02$ and $r_- = 0.29 \pm 0.04$, obtained when $|p_+|$ and $|p_-|$
were not assumed to be equal. (Somewhat smaller values, $r_- = 0.25 \pm 0.03$, were 
obtained in the global fit when assuming $p_+ = -p_-$, 
as proposed in~\cite{Lipkin}.) 
Values on the low side, $r_+ = 0.10^{+0.06}_{-0.04}$ and $r_- = 0.10^{+0.09}_{-0.05}$,
were calculated in QCD factorization~\cite{BN}. 

Assuming ${\cal A}^{\pm}_{\rm CP} \approx 0$, Eqs.~(\ref{alphaR+}) and (\ref{alphaR0}) 
imply
\beq\label{boundsalpha+-}
|\sin(\alpha^+_{\rm eff} - \alpha)| \le \sqrt{{\cal R}^0_+}~,~~~~
|\sin(\alpha^+_{\rm eff} - \alpha)| \le \sqrt{{\cal R}^+_+}|\sin(\beta + \alpha)|~,~~~~
|\sin(\alpha^-_{\rm eff} - \alpha)| \le \sqrt{{\cal R}^0_-}~.
\eeq
Note that in (\ref{alphaR0}) ${\cal A}^-_{\rm CP}$ is $3.3 \sigma$ away from zero;
nonzero asymmetries would improve the above bounds.
Currently we find the following upper limits at $90\%$ 
confidence level,
\beq\label{alphabounds}
{\cal R}^0_+\Rightarrow~|\alpha^+_{\rm eff} - \alpha| \le 14.2^\circ~,~~~~
{\cal R}^+_+\Rightarrow~|\alpha^+_{\rm eff} - \alpha| \le 7.3^\circ - 11.7^\circ~,~~~~
{\cal R}^0_-\Rightarrow~|\alpha^-_{\rm eff} - \alpha| \le 15.4^\circ~,
\eeq
where the central upper limit (obtained by using $38^\circ \le \gamma \le 80^\circ$)
is shown to improve slightly as $\alpha$ increases in 
the range $78^\circ \le \alpha \le 122^\circ$~\cite{CKMfitter}. Using (\ref{alphaeff}),
the second and third upper limits imply
\beq\label{Delta-alpha}
|\alpha_{\rm eff} - \alpha| \le 11.3^\circ - 13.5^\circ~,
\eeq
where the bound is improved slightly as $\alpha$ becomes larger within the above range.
A recent study of $\alpha$ in time-dependent asymmetries in 
$B^0\to\pi^+\pi^-$~\cite{pipi} favors the upper part of this range.

The upper bound (\ref{Delta-alpha}), which may be improved by a few degrees 
through more precise measurements of branching ratios, 
including a first observation of $B^+\to \rho^+K^0$,
is expected to be modified 
by SU(3) breaking. For instance, if SU(3) breaking enhances $p'_{\pm}$ relative to 
$p_{\pm}$ (as it would by assuming factorization for these amplitudes), then the upper 
bound becomes stronger. In this respect these upper limits may be considered 
conservative. In any event, even if SU(3) breaking suppresses $p'_{\pm}$ relative to 
$p_{\pm}$ by 20 or 30 percent, one expects the upper bounds to change by this 
amount. This result provides an important conclusion, implying  that 
{\em in time-dependent decays $B^0\to \rho^{\pm}\pi^{\mp}$  $\alpha$ 
may be measured through $\alpha_{\rm eff}$ with a precision of about 
$\pm 15^\circ$}:
\beq\label{alphaeff-alpha}
|\alpha_{\rm eff} - \alpha| \le 15^\circ~~({\rm including~SU(3)~breaking})~.
\eeq
This accuracy is comparable to that of measuring $\alpha$ through $\alpha_{\rm eff}$ in 
time-dependent $B^0\to\rho^+\rho^-$ decays~\cite{rho+rho-}.
Here the shift caused by the penguin amplitude is constrained in
the isospin analysis~\cite{Gronau:1990ka,GQ,FLNQ} by measured branching
ratios of $B^0\to\rho^+\rho^-, B^+\to\rho^+\rho^0$ and by an upper bound on
$B^0\to \rho^0\rho^0$~\cite{rho0rho0} to a range,  
$|\alpha_{\rm eff} - \alpha| \le 17^\circ$, at 90\% confidence level. 

Using Eq.~\eqref{alphaeff}, the current data (\ref{CSDelta}) may now be translated
into solutions for $\alpha_{\rm eff}$. 
In order to reduce discrete ambiguities caused by the few branches of the two 
$\arcsin$ functions in (\ref{alphaeff}), we note that 
\beq
(2\alpha^+_{\rm eff} +\hat\delta) - (2\alpha^-_{\rm eff} - \hat\delta) = 2\delta_t + 
{\cal O}(r_{\pm})~.
\eeq
We will make a conservative assumption that the two angles on the left hand side 
differ by much less than $180^\circ$, which is equivalent to assuming that 
$\delta_t$ is much smaller than $90^\circ$. (Note that QCD factorization predicts a
very small value, $\delta_t = (1\pm 3)^\circ$~\cite{BN}, while a global SU(3) fit 
finds $\delta_t \simeq -20^\circ$~\cite{CGLRS}.)
This mild assumption can be checked experimentally by measuring the phase difference 
$\arg(A_-/A_+)$ using the overlap of the $\rho^+$ and $\rho^-$ resonance bands in 
the $B^0 \to\pi^+\pi^-\pi^0$ Dalitz plot \cite{Aubert:2004iu}. (This measurement is expected to be feasible
much before a complete isospin and Dalitz plot analysis~\cite{Snyder:1993mx}
can be performed.)~The measurable phase difference $\arg(A_-/A_+)$
is dominated by $\delta_t = \arg(t_-/t_+)$.
The difference, $|{\rm arg}(A_-/A_+)-\delta_t|$, is governed by the subdominant 
amplitudes $p_{\pm}$. We have checked that this difference is less then $25^\circ$ 
at 90$\%$ confidence level, when $r_+$ is in the range \eqref{r+R++}, consistent with 
(\ref{r+-}) and (\ref{expR}), and when $r_-$ is in the range \eqref{r-bound}, consistent with
(\ref{r+-}) where $R^0_-$ is replaced by 
$\Gamma(B^0\to\rho^-K^+)/\Gamma(B^0\to \rho^-\pi^+)$.
Therefore, a small measured value of $\arg(A_-/A_+)$ would 
imply that $\delta_t$ is much smaller than $90^\circ$, confirming our assumption. 

Applying (\ref{alphaeff}) separately to the BABAR and BELLE measurements, (which differ 
in their $\Delta S$ values by $2\sigma$) and to their averages,
we find that in all three cases the experimental errors in $C,\Delta C, S$ and $\Delta S$
translate into quite small errors in $\alpha_{\rm eff}$, $\pm 5^\circ,  \pm 13^\circ$ and 
$\pm 4^\circ$ respectively. 
In each case one finds for the central values of these 
measurements only two solutions in the range $0 \le \alpha_{\rm eff} \le \pi$:
\beq\label{alphalist}
\alpha_{\rm eff} =  \left\{ \begin{array}{ccccccccc}
{\rm BABAR}: & 93^\circ, & 177^\circ \cr
{\rm BELLE}: & 102^\circ,& 168^\circ\cr
{\rm Average}: & 94^\circ, & 175^\circ\cr
 \end{array} \right.~.
\eeq
Excluding by $\alpha+\beta < \pi$ the three values near $180^\circ$, corresponding to an ambiguity 
$\alpha \to 3 \pi/2 -\alpha$, we find
\beq\label{alphaexp}
\alpha =  \left\{ \begin{array}{ccccccccc}
(93 \pm 5 \pm 15)^\circ & {\rm BABAR} \cr 
(102 \pm 13 \pm 15)^\circ & {\rm BELLE} \cr 
(94 \pm 4 \pm 15)^\circ & {\rm Average} \cr 
 \end{array} \right.~,
\eeq
where the first error is experimental and the second is theoretical, coming from the bound \eqref{alphaeff-alpha}. Note the weak dependence of $\alpha$ on $\Delta S$, the 
central values of which have opposite signs in the BABAR and BELLE 
measurements~(\ref{CSDelta}). Combining for simplicity the experimental  and 
theoretical errors~\eqref{alphaexp} in quadrature gives for the average
\beq
\alpha=(94\pm 16)^\circ~.
\eeq
All the above results are in good agreement with the range $78^\circ \le \alpha \le 122^\circ$
obtained from other CKM constraints~\cite{CKMfitter}. 
For comparison, we note that the world average CP asymmetries measured in 
$B^0\to \pi^+\pi^-$~\cite{Jawahery:2003,Bellepipi} have been recently studied in Ref.~\cite{pipi} 
and were shown to imply a comparable range, $\alpha = (103 \pm 17)^\circ$, favoring large
values of the weak phase in this range.

\section{Extracting $\alpha$}\label{extracting_alpha}
As mentioned, the observables in $B\to \rho^\pm \pi^\mp$ and the SU(3)
related rates are sufficient for determining $\alpha$. One has four independent 
observables $\Sbar$, $\Delta \Sbar$, 
${\cal A}_{CP}^{\pm}$ in time-dependent decays $B\to \rho^\pm \pi^\mp$, and additional 
four independent observables ${\cal R}_\pm^0$, ${\cal R}_\pm^+$ in $\Delta S=1$ 
decays (where currently only an upper bound on ${\cal R}_-^+$ exists). These eight observables 
provide an overdetermined set of conditions, as they depend on only six parameters, the hadronic parameters $r_\pm$, $\delta_\pm$, $\delta_t$ and the weak angle $\alpha$. 
The set of Eqs. \eqref{ACP+-}, \eqref{Sbar-full}, \eqref{DeltaSbar-full} and \eqref{R*0} then allows 
for an extraction of $\alpha$ as well as all the hadronic parameters. No assumption is 
required about $\delta_t$, thus relaxing the mild 
and experimentally testable assumption made in the
previous section  in order to obtain the unambiguous ranges (\ref{alphaexp}).

\begin{figure}
\begin{center}
\epsfig{file=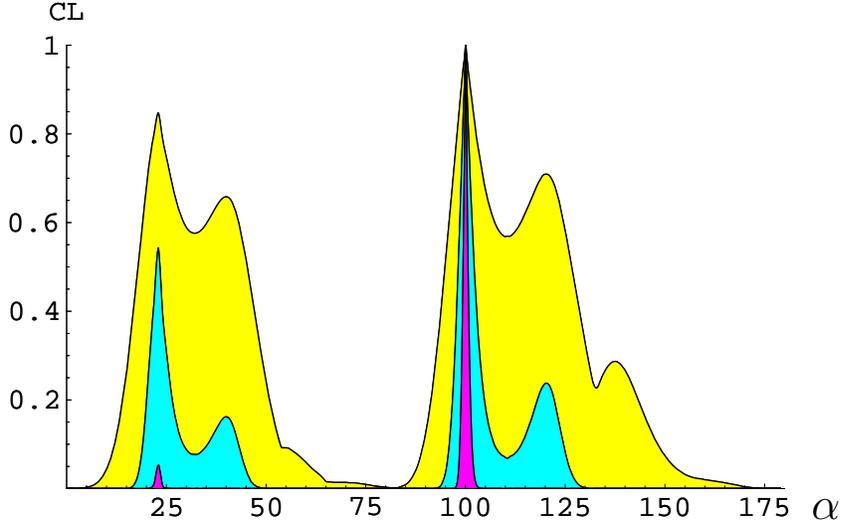, height=7.0cm}
\rput*(-0.1,0.2){{\Large $\alpha$}}
\caption{\footnotesize{Confidence level (CL) as a function of $\alpha$ for a generated
set of data using a choice of parameters, $r_+=0.18$,
$r_-=0.23$, $\delta_+=30^\circ$, $\delta_-=-55 ^\circ$,
$\delta_t=170^\circ$ and $\alpha_{\rm input}=100^\circ$. 
The interpretation of a confidence level is the same as in 
Ref.~\cite{CKMfitter} when Gaussian errors are assumed.
Errors used for
$\chi^2$ are the currently measured ones [yellow (light gray) region], those anticipated with 
ten times statistics [cyan (gray)] and hundred times statistics [purple (dark gray)]. We assume an experimental
error in ${\cal R}_-^+$ as in ${\cal R}_-^0$.}} \label{FigPValue}
\end{center}
\end{figure}

A solution of Eqs. \eqref{ACP+-}, \eqref{Sbar-full},
\eqref{DeltaSbar-full} and \eqref{R*0} under these general conditions
is shown for illustration in Fig. \ref{FigPValue}, which plots  
confidence levels (CL) as functions of $\alpha$ for different levels of statistics. 
To obtain the plots we generated data for the observables  $\Sbar$, $\Delta 
\Sbar$, ${\cal A}_{CP}^{\pm}$, ${\cal R}_\pm^0$ and ${\cal R}_\pm^+$, using 
a particular choice of values for the parameters $\delta_\pm$, $r_\pm$, $\delta_t$ 
(as specified in the figure caption) and an input value $\alpha=100^\circ$.
The errors on the observables were taken to be
the currently measured ones, apart from an error on ${\cal R}_-^+$, for which 
the error was taken to be the same as the current error on ${\cal R}_-^0$. 
Improvements in confidence level are shown for ten and hundred times more 
data than available today. One sees that with enough statistics only one solution 
at $\alpha = 100^\circ$ survives in the range $0^\circ \le \alpha \le180^\circ$. 
That is, for this particular choice of parameters one hundred times the present statistics implies an uncertainty of 
$\pm  2^\circ$  in the single value of $\alpha$ extracted at 95 \% CL. We checked that the situation presented in Fig. 
\ref{FigPValue} is generic and applies to a large range of hadronic parameters.

The ambiguities in $\alpha$, which are eventually resolved with 
high enough statistics, are seen in Fig.~2 to imply
a large range of allowed values of $\alpha$ at current
statistics. To get a quick insight into the origin of these ambiguities, 
let us first explore the case of $r_\pm=0$. In this (oversimplified) case,
involving merely mixing induced CP violation,  the only observables 
carrying information on $\alpha$ are $\Sbar$ and $\Delta \Sbar$,
\beq
\begin{split}\label{r=0}
\Sbar&=\sin2\alpha \cos\delta_t~,\\
\Delta\Sbar&=\cos 2\alpha  \sin\delta_t~.
\end{split}
\eeq
Assuming that $\Sbar$ and $\Delta \Sbar$ are measured precisely,
a solution for $\sin 2\alpha$ is given by
\beq
\big(\sin 2\alpha\big)^2=\frac{1}{2} \left\{ 1+\Sbar^2- (\Delta
\Sbar)^2\pm \sqrt{\big[1 +  \Sbar^2-(\Delta \Sbar)^2\big]^2-4
\Sbar^2}\right\}~.\label{analytic}
\eeq
The two signs in front of the square root of the discriminant
correspond to the following map 
\beq
P_{t}=\{\alpha\to \frac{\pi}{4}-\frac{\delta_t}{2}, \quad \delta_t\to
\frac{\pi}{2} -2\alpha\}~,\label{PtLO}
\eeq
or equivalently to the following interchange in Eqs.~\eqref{r=0}:
\beq
\begin{split}
\sin 2\alpha&\leftrightarrow \cos\delta_t~,\\
\cos 2\alpha &\leftrightarrow \sin\delta_t~.
\end{split}
\eeq
The other discrete ambiguities of \eqref{r=0} are
\begin{align}\label{Ppi}
P_\pi&=\{\alpha\to\alpha +\pi\}~,\\
\label{Ppi/2}
P_{\pi/2}&=\{\alpha\to \frac{\pi}{2}-\alpha, \quad \delta_t\to
-\delta_t\}~,\\
\label{P-}
P_-&=\{\alpha\to -\alpha, \quad \delta_t\to \pi-\delta_t\}~.
\end{align}

While none of these transformations relate to one another two values 
of $\alpha$ in the allowed range, $78^\circ \le \alpha \le 
122^\circ$~\cite{CKMfitter}, combinations of these transformations, such as 
$P_\pi P_{\pi/2} = \{\alpha\to 3\pi/2 -\alpha,...\}$, are relevant to  this range.
However, this ambiguity, as well as the others, is resolved once
higher order terms in $r_\pm$ are taken into account, as can be seen in 
Fig.~2. (See Appendix A for
details.) Namely, the complete set of equations, \eqref{ACP+-}, 
\eqref{Sbar-full}, \eqref{DeltaSbar-full} and \eqref{R*0},  is not invariant 
under these  transformations, which are violated by terms of order $r_\pm$. 
Note that although the ratios ${\cal R}^{0,+}_{\pm}$ are formally of order $r^2_{\pm}$
(because of the multiplicative factor $\lambdaB^2$ in their definitions), 
they are in fact zeroth order. That is, they must be measured to an accuracy of 
order $r_{\pm}$  in order to resolve the ambiguities. It turns out that at least one 
pair of $(R^0_{\pm}, R^+_{\pm})$ ought to be measured to this precision. 
(See Appendix A.)

Since the extraction of $\alpha$ relies on a given scheme of broken flavor  
SU(3), one may wonder how SU(3) breaking effects other than those included 
may affect the value of $\alpha$. Flavor SU(3) is used to fix the values of
$r_\pm$, which we have shown to be small, $r_{\pm}\sim 0.2$. The extracted 
values of $\alpha$ are given to zeroth order in $r_{\pm}$ by 
\eqref{analytic}. Terms of order $r_\pm$ are affected by SU(3) breaking corrections, which by themselves are approximately
$r_\pm$. Therefore the overall SU(3) breaking effect in $\alpha$ is expected to be of order $r_\pm^2$.

This is demonstrated in the following way.
We first generate values for observables by randomly varying $\delta_t, \delta_\pm$
and $\alpha$, and taking $r_\pm$ to vary in the allowed ranges \eqref{r+R++}
and \eqref{r-bound}. 
In addition to SU(3) breaking in tree amplitudes we now allow also for the SU(3) breaking
in penguin amplitudes by introducing positive parameters, $0.7<c_\pm<1.3$, and writing
$p'_{\pm} = -c_{\pm}\lambdaB^{-1}p_{\pm}$ instead of \eqref{t'+}. 
[We neglect SU(3) breaking in $\delta_{\pm}$ which is nonleading in the sense that the 
term $2r_{\pm}\lambdaB^2_{\pm}z_{\pm}$ in the numerator of the first of Eq.~(\ref{R*0}) 
is smaller than the $r^2_{\pm}$ term.] Generating data under this 
assumption, we then extract $\alpha$ using Eqs.~\eqref{ACP+-}, \eqref{Sbar-full}, 
\eqref{DeltaSbar-full}
and \eqref{R*0},  where SU(3) was assumed to be present only in tree amplitudes.
Running a Monte Carlo program for 10000 different configurations shows that the local minimum 
in $\chi^2$ shifts by only $\sqrt{\langle(\alpha^{\rm out}-\alpha^{\rm in})^2\rangle}\sim 2^\circ$,
which is indeed of order $r^2_{\pm}$. 

There exist experimental tests of flavor SU(3) and SU(3) breaking corrections, in terms of 
equalities between CP rate differences in SU(3) related processes. Two of these relations 
follow from Eqs.~(\ref{Apm}) and (\ref{AmpTreePeng'})~\cite{CGLRS},
\beqa\label{DeltaGamma1}
\Gamma(B^0\to\rho^+\pi^-) - \Gamma(\Bbar \to \rho^-\pi^+) & = &
\frac{f_{\rho}}{f_{K^*}}\left [ \Gamma(\Bbar \to K^{*-}\pi^+) - \Gamma(B^0\to K^{*+}\pi^-)\right ]~,
\\
\label{DeltaGamma2}
\Gamma(B^0\to\rho^-\pi^+) - \Gamma(\Bbar \to \rho^+\pi^-) & = &
\frac{f_{\pi}}{f_K}\left [ \Gamma(\Bbar \to \rho^+ K^-) - \Gamma(B^0\to \rho^-K^+)\right ]~.
\eeqa
These relations test the equality of products $|t^{(')}_\pm|
|p^{(')}_\pm|\sin\gamma\sin\delta_\pm$ in $\Delta S=0$
and $\Delta S=1$ decays. Using current values in Table \ref{table-1}, 
one finds that,
while the signs of the two asymmetries in the second equality confirm the SU(3) 
prediction, their absolute values differ by  $2.0\sigma$.


\section{Conclusion}
We have studied implications for $\alpha$ of time-dependent rate measurements in 
$B\to\rho^{\pm}\pi^{\mp}$ by proposing a parametrization which depends on 
a minimal number of hadronic parameters and observables. We have proposed one 
method based on factorization, which reduces by two the number of parameters to 
the number of observables. The limitation of this method was shown to be its sensitivity 
to terms quadratic in $r_{\pm}$, the two ratios of penguin and tree amplitudes. 

Assuming broken flavor SU(3), which relates $B\to\rho^{\pm}\pi^{\mp}$ to four processes 
of the form $B^{0,+}\to K^*\pi$ and $B^{0,+}\to \rho K$, and using branching 
ratios measured for these processes, we calculated lower and upper bounds 
on $r_{\pm}$, slightly below and slightly above 0.2. Defining a measurable 
quantity $\alpha_{\rm eff}$, that becomes $\alpha$ in the limit of vanishing penguin 
amplitudes, we calculated upper bounds on $|\alpha_{\rm eff} -\alpha|$ 
in a range $11^\circ - 13^\circ$, which are expected to be at most about 
$15^\circ$ when including unaccounted SU(3) breaking effects. 

In order to resolve a discrete ambiguity in $\alpha$, we assumed that the relative strong 
phase of two tree amplitudes, $\delta_t$, is considerably smaller than $90^\circ$. 
This assumption, justified by QCD factorization and by a global SU(3) fit
to $B\to VP$ decays, can be tested directly through a partial Dalitz plot analysis
of $B\to \pi^+\pi^-\pi^0$.
Using the BABAR and BELLE results for $B(t) \to \rho^{\pm}\pi^{\mp}$
this then implies single solutions, $\alpha = (93\pm 16)^\circ$ and $(102 \pm 20)^\circ$,
respectively, and an average $\alpha = (94 \pm 16)^\circ$, taking into account an error 
scaling factor. 

Finally, using a complete set of measurables, including CP asymmetries 
and avoiding any assumption about $\delta_t$, we presented numerical studies 
demonstrating the feasibility of determining $\alpha$ and the reduction of discrete 
ambiguities with statistics. We have also shown that SU(3) breaking effects, which were 
not already included, are expected to be very small, of order $r^2_{\pm}$.

\section*{Acknowledgments}
We wish to thank Tom Browder for informing us about the BELLE time-dependent 
measurements of $B(t)\to\rho^{\pm}\pi^{\mp}$ and for motivating this study.
We also thank Jonathan Rosner and Denis Suprun for useful comments.
The work of J. Z. is supported in part by EU grant HPRN-CT-2002-00277 and  
by the Ministry of Education, Science and Sport of the Republic of Slovenia.

\appendix
\section{Further details on discrete ambiguities}
Discrete ambiguities were found in Section VII  when order $r_\pm$ corrections to
observables were neglected. This lead to a 16-fold ambiguity on
$\alpha$ in the range $\alpha \in [0,2 \pi)$ spanned by the
transformations $P_t$, $P_\pi$, $P_{\pi/2}$ and $P_-$ given in
\eqref{PtLO}, \eqref{Ppi}, \eqref{Ppi/2} and \eqref{P-}. Let us now discuss how the
higher order terms in $r_\pm$ affect the ambiguities. First of all, the 
(unphysical) transformation $P_\pi$ is  an exact symmetry of
Eqs. \eqref{ACP+-}, \eqref{Sbar-full}, \eqref{DeltaSbar-full}, and
\eqref{R*0}, if extended to a transformation on strong phases
\beq
P_\pi=\{\alpha\to \alpha+\pi, \delta_\pm\to \delta_\pm+\pi, \delta_t\to
\delta_t\}~. \label{P_pi}
\eeq

The other symmetry transformations  $P_i=\{P_t,P_{\pi/2}, P_-\}$ receive 
higher order corrections. To see under which conditions
they remain ambiguities, let us expand Eqs.~\eqref{ACP+-}, \eqref{Sbar-full}, 
\eqref{DeltaSbar-full}, and (\ref{R*0}) to first order in $r_\pm$, where we count 
$\delta_t\sim r_\pm\sim \bar{\lambda}\ll 1$,
\begin{align}
\begin{split}
\Sbar-\sin 2\alpha =~&\sin2\alpha \left[r_+\cos(\beta+\alpha+\delta_+)+r_-\cos(\beta+\alpha+\delta_-)\right]\\
&-\sin(
\alpha-\beta)\left[r_+\cos\delta_++r_-\cos\delta_-\right]~,
\end{split}\label{SbarCorrection}
\\
\Delta \Sbar =~&\cos 2 \alpha
\sin\delta_t+(r_+\sin\delta_+-r_-\sin\delta_-) \cos(\alpha -\beta)~,\label{DeltaSbarCorrection}
\\
\label{ACP+-lin}
{\cal A}^{\pm}_{\rm CP} =~& -2r_{\pm}\sin\delta_{\pm}\sin(\beta + \alpha)~,\\
\frac{1}{2}\left({\cal R}_\pm^0-{\cal R}_\pm^+\right) =~&r_\pm
\bar{\lambda}_\pm^2 \cos(\delta_\pm)\cos(\beta+\alpha)~.\label{diffR}
\end{align}

The parameters $r_\pm$ are obtained from 
\beq
\sqrt{{\cal R}_\pm^0}=\sqrt{{\cal R}_\pm^+}=r_\pm~.\label{1+rend}
\eeq
One is then left with six
Eqs. \eqref{SbarCorrection}, \eqref{DeltaSbarCorrection},
\eqref{ACP+-lin} and (\ref{diffR}) for four unknowns, $\delta_\pm$, $\delta_t$, and
$\alpha$.  
In order to check for leftover ambiguities, let us assume that 
there exists at least one solution for \eqref{SbarCorrection}, 
\eqref{DeltaSbarCorrection},
\eqref{ACP+-lin}, and (\ref{diffR}) which we denote by $\delta_\pm^0$, $\delta_t^0$, and
$\alpha^0$. The transformations $P_i$ may give us another viable
solution for $\alpha$,
\beq
\alpha^{\rm new}=P_i \alpha^0 +\delta\alpha~, \label{alphanew}
\eeq
together with new  values for the other paremeters, $\delta_\pm^{\rm
new}$, $\delta_t^{\rm new}$. From \eqref{ACP+-lin} we get (to leading
order in the small parameters, $r_\pm, \delta_t^{\rm new}$ and $\delta\alpha$),
\beq
\sin\delta_\pm^{\rm new}=-\frac{ {\cal A}^{\pm}_{\rm CP}}{
2r_{\pm}\sin(\beta + P_i\alpha^0)}~, \label{sindeltapm}
\eeq
while  \eqref{diffR} gives
\beq\label{cosdeltapm}
\cos\delta_\pm^{\rm new}=\frac{1}{2}\frac{\left({\cal R}_\pm^0-{\cal 
R}_\pm^+\right)}{r_\pm \bar{\lambda}_\pm^2\cos(\beta+P_i\alpha^0)}~.
\eeq
In general Eqs. \eqref{sindeltapm}, \eqref{cosdeltapm} are not simultaneously satisfied 
which resolves the ambiguity. 

In case that  ${\cal R}_\pm^{0,+}$  is not measured to order $r_{\pm}$, i.e. to
a precision of about $20\%$, the ambiguity is retained if the 
right hand side of (\ref{sindeltapm}) is not larger 
in magnitudes than one, leading to a solution for 
\eqref{SbarCorrection} and \eqref{DeltaSbarCorrection}: 
\begin{align}
\delta_t^{\rm new}=&\frac{1}{\cos (2 P_i\alpha^0)}\left[\Delta
\Sbar-(r_+\sin\delta_+^{\rm new}-r_-\sin\delta_-^{\rm new})
\cos(P_i\alpha^0 -\beta)\right]~,\\
\begin{split}
\delta\alpha=&\frac{1}{2\cos(2 P_i \alpha^0)}\Big\{\Sbar-\sin 2\alpha^0
+\sin(
P_i\alpha^0-\beta)\left[r_+\cos\delta_+^{\rm new}+r_-\cos\delta_-^{\rm new}\right] \\
&-\sin2\alpha^0 \left[r_+\cos(\beta+P_i\alpha^0+\delta_+^{\rm
new})+r_-\cos(\beta+P_i\alpha^0+\delta_-^{\rm new})\right]\Big\}~. \label{deltaalpha}
\end{split}
\end{align}
These expressions show the existence of further ambiguities in $\delta\alpha$ 
of order $r_{\pm}$ caused by twofold solutions for $\delta_{\pm}^{\rm new}$
in  (\ref{sindeltapm}) or (\ref{cosdeltapm}).
[Note that $\delta_t^{\rm new}$ and $\delta\alpha$ are
$O(r_\pm)$ in accordance with our expansion.]
This shows the importance of measuring ${\cal R}_\pm^{0,+}$ as precisely 
as possible.


\def \ajp#1#2#3{Am.\ J. Phys.\ {\bf#1}, #2 (#3)}
\def \apny#1#2#3{Ann.\ Phys.\ (N.Y.) {\bf#1}, #2 (#3)}
\def \app#1#2#3{Acta Phys.\ Polonica {\bf#1}, #2 (#3)}
\def \arnps#1#2#3{Ann.\ Rev.\ Nucl.\ Part.\ Sci.\ {\bf#1}, #2 (#3)}
\def \art{and references therein}
\def \cmts#1#2#3{Comments on Nucl.\ Part.\ Phys.\ {\bf#1}, #2 (#3)}
\def \cn{Collaboration}
\def \cp89{{\it CP Violation,} edited by C. Jarlskog (World Scientific,
Singapore, 1989)}
\def \econf#1#2#3{Electronic Conference Proceedings {\bf#1}, #2 (#3)}
\def \efi{Enrico Fermi Institute Report No.}
\def \epjc#1#2#3{Eur.\ Phys.\ J.\ C {\bf#1}, #2 (#3)}
\def \ib{{\it ibid.}~}
\def \ibj#1#2#3{~{\bf#1} (#3) #2}
\def \ijmpa#1#2#3{Int.\ J.\ Mod.\ Phys.\ A {\bf#1}, #2 (#3)}
\def \ite{{\it et al.}}
\def \jhep#1#2#3{JHEP {\bf#1}, #2 (#3)}
\def \jpb#1#2#3{J.\ Phys.\ B {\bf#1}, #2 (#3)}
\def \kdvs#1#2#3{{Kong.\ Danske Vid.\ Selsk., Matt-fys.\ Medd.} {\bf #1}, No.\
#2 (#3)}
\def \mpla#1#2#3{Mod.\ Phys.\ Lett.\ A {\bf#1}, #2 (#3)}
\def \nat#1#2#3{Nature {\bf#1}, #2 (#3)}
\def \nc#1#2#3{Nuovo Cim.\ {\bf#1}, #2 (#3)}
\def \nima#1#2#3{Nucl.\ Instr.\ Meth.\ A {\bf#1}, #2 (#3)}
\def \npb#1#2#3{Nucl.\ Phys.\ B~{\bf#1},  #2 (#3)}
\def \npps#1#2#3{Nucl.\ Phys.\ Proc.\ Suppl.\ {\bf#1}, #2 (#3)}
\def \PDG{Particle Data Group, D. E. Groom \ite, \epjc{15}{1}{2000}}
\def \pisma#1#2#3#4{Pis'ma Zh.\ Eksp.\ Teor.\ Fiz.\ {\bf#1}, #2 (#3) [JETP
Lett.\ {\bf#1}, #4 (#3)]}
\def \pl#1#2#3{Phys.\ Lett.\ {\bf#1}, #2 (#3)}
\def \pla#1#2#3{Phys.\ Lett.\ A {\bf#1}, #2 (#3)}
\def \plb#1#2#3{Phys.\ Lett.\ B {\bf#1}, #2 (#3)} 
\def \prd#1#2#3{Phys.\ Rev.\ D\ {\bf#1}, #2 (#3)}
\def \prl#1#2#3{Phys.\ Rev.\ Lett.\ {\bf#1}, #2 (#3)} 
\def \prp#1#2#3{Phys.\ Rep.\ {\bf#1}, #2 (#3)}
\def \ptp#1#2#3{Prog.\ Theor.\ Phys.\ {\bf#1}, #2 (#3)}
\def \rmp#1#2#3{Rev.\ Mod.\ Phys.\ {\bf#1}, #2 (#3)}
\def \rp#1{~~~~~\ldots\ldots{\rm rp~}{#1}~~~~~}
\def \yaf#1#2#3#4{Yad.\ Fiz.\ {\bf#1}, #2 (#3) [Sov.\ J.\ Nucl.\ Phys.\
{\bf #1}, #4 (#3)]}
\def \zhetf#1#2#3#4#5#6{Zh.\ Eksp.\ Teor.\ Fiz.\ {\bf #1}, #2 (#3) [Sov.\
Phys.\ - JETP {\bf #4}, #5 (#6)]}
\def \zp#1#2#3{Zeit.\ Phys.\  {\bf#1}, #2 (#3)}
\def \zpc#1#2#3{Zeit.\ Phys.\ C {\bf#1}, #2 (#3)}
\def \zpd#1#2#3{Zeit.\ Phys.\ D {\bf#1}, #2 (#3)}


\end{document}